\documentclass[lineno]{jfm}
\usepackage{graphicx}
\usepackage{epstopdf,epsfig}
\usepackage{newtxtext}
\usepackage{newtxmath}
\usepackage{natbib}
\usepackage{hyperref}
\usepackage{xcolor}
\usepackage{amsmath}
\usepackage{amsfonts}
\usepackage{pgfplots}
\usepackage{breqn}
\pgfplotsset{compat=1.18}
\usepackage{algpseudocode}
\usepackage{algorithm}
\usepackage{varwidth}
\usepackage{float}
\usepackage{academicons}
\usepackage{tikz}
\usepackage{siunitx}
\usepackage{graphicx} 
\usepackage{multirow} 
\usepackage{comment} 
\usepackage{pgfplots} 
\usepackage{adjustbox}  
\usepackage{xcolor}    
\usepackage{ragged2e} 
\usepackage{multicol} 
\usepackage{rotating}
\usepackage{caption}
\usepackage{subcaption}
\usepackage{booktabs}
\usepackage{nth}
\usepackage{textpos}
\usepackage{braket} 
\usepackage{orcidlink}

\usepackage{tikz}
\usetikzlibrary{arrows.meta, positioning}
\usepackage[dvipsnames]{xcolor}
\usepackage{pgfplots}
\pgfplotsset{compat=newest}
\usetikzlibrary{backgrounds}
\usepackage{float}
\usepackage{amsmath}
\usepackage{amssymb}
\usepackage{algorithm}
\usepackage{algpseudocode}
\usepackage{cleveref}
\usetikzlibrary{arrows}
\usetikzlibrary{arrows.meta}
\usetikzlibrary{fit,positioning}
\usetikzlibrary{shapes.geometric}
\definecolor{orcidlogocol}{HTML}{A6CE39}
\hypersetup{
    colorlinks = true,
    urlcolor   = blue,
    citecolor  = black,
}

\newcommand{\RomanNumeralCaps}[1]
\linenumbers
\title{Latent-Space Non-Linear Model Predictive Control for Partially-Observable Systems}
\author{
Luigi Marra\aff{1} \corresp{\email{luigi.marra@uc3m.es}},
Onofrio Semeraro\aff{2},
Lionel Mathelin\aff{2},
Andrea Meilán-Vila\aff{3},
\and 
Stefano Discetti\aff{1}}

\affiliation{
\aff{1}
Department of Aerospace Engineering, Universidad Carlos III de Madrid, Av. de la Universidad 30, Leganés, 28911, Madrid, Spain
\aff{2}
CNRS, Laboratoire interdisciplinaire des sciences du numérique (LISN), Université Paris-Saclay, 91400 Orsay, France
\aff{3}
Department of Statistics, Universidad Carlos III de Madrid, Av. de la Universidad 30, Leganés, 28911, Madrid, Spain}
\shortauthor{
L. Marra, 
O. Semeraro, 
L. Mathelin,
A. Meilán-Vila,  
S. Discetti }

\begin{document}
\maketitle
\begin{abstract}\nolinenumbers This work presents a scalable control framework based on nonlinear Model Predictive Control for high-dimensional dynamical systems. 
The proposed approach addresses the key challenges of model scalability and partial observability by integrating data-driven reduced order modelling, control in a latent space, and state estimation within a unified formulation. A predictive model is constructed via Operator Inference on a Proper Orthogonal Decomposition basis, yielding a compact latent representation that captures the dominant system dynamics. State estimation is achieved through an Unscented Kalman Filter, which reconstructs the latent space from sparse and noisy measurements, enabling closed-loop control. The input signals are computed directly in the reduced-order latent space, improving computational efficiency with negligible impact on predictive capability. The methodology is validated on the one- and two-dimensional Kuramoto–Sivashinsky equations, serving as benchmarks for chaotic and spatially-extended systems. Numerical experiments demonstrate that the proposed framework achieves accurate stabilisation. Overall, the framework provides a practical approach for nonlinear control of complex, high-dimensional systems where full-state measurements are often inaccessible or infeasible.
\end{abstract}
\nolinenumbers
\section{Introduction}\nolinenumbers
\label{app:introduction}

%%%%%% The challenges in control 
Controlling dynamical systems arising in scientific, engineering, and industrial applications can pose significant challenges, as these systems may exhibit complex behaviours.
Such systems often exhibit high dimensionality, multiscale spatio-temporal dynamics, strong nonlinearities, time delays, or chaotic behaviour \citep{Brunton_data_driven_2022}.
These characteristics make the design of efficient real-time control strategies both computationally demanding and, in many cases, practically infeasible. Such difficulties are particularly evident in fluid dynamics, where flows are typically two- or three-dimensional, governed by nonlinear equations, and characterised by intricate multiscale structure \citep{Brunton_closedloop_turbulence_2015}.

%%%%%% Model-based strategies
A common strategy is model-based control, in which a possibly surrogate model of the system dynamics is used to determine control actions. These approaches, implicitly or explicitly, incorporate knowledge of the system, sometimes using a predictive model to simulate the response under different inputs. However, for complex nonlinear systems, accurate long-term prediction is rarely possible. Sensitivity to initial conditions, unmodeled dynamics, and measurement noise typically restrict reliable predictions to short horizons.

%%%%%% Model predictive control and challenges
Model Predictive Control \citep[MPC,][]{Camacho_mpc_2007} is a powerful control framework that uses predictive models over a finite horizon to determine control actions. At each time step, MPC solves an optimisation problem of a prescribed cost functional, which encodes the control objectives. This produces a sequence of future control actions, of which only the first is implemented. The horizon then shifts forward, and the optimisation is repeated using updated system information. This strategy, known as receding-horizon control, gives MPC both flexibility and robustness, enabling its application to linear \citep{Muske_linear_mpc_1993} and nonlinear systems \citep{allgower_nmpc_2004}, as well as to continuous and discrete-time formulations. By directly optimising over model predictions, MPC also allows for the incorporation of input and state constraints, making it suitable for safety-critical applications even within fluid dynamics. In this field, MPC has been successfully applied to a variety of problems, such as wall-bounded flows \citep{Bewley_dns_mpc_2001}, cavity flows \citep{Arbabi_koopman_mpc_2018}, and wake flows such as those past circular cylinders \citep{Morton_deep_mpc_2018,Deda_NN_mpc_2023} and the fluidic pinball \citep{Marra_SelfTuningMpc_2024, Bieker_deep_mpc_2020}. A comparison of MPC with other model-based techniques for flow control is reported in \cite{Fabbiane_mpc_unstable_flows_2014}. The cornerstone of the application of MPC is the availability of a reliable predictive model of the system dynamics that is accurate, computationally efficient (potentially parsimonious), robust, and whose states are directly measurable. However, for high-dimensional nonlinear dynamical systems, achieving all these properties poses a major challenge.

Regarding model construction, the choice of the approach depends on the availability of governing equations and data. In many practical settings, the equations are known and high-fidelity simulations are available, but their computational cost makes them unsuitable for direct use in control. In such cases, reduced-order modelling (ROM) provides an efficient way to approximate the essential system dynamics. These methods employ a proxy for the full system state, significantly improving computational efficiency in control. Projection-based approaches, for example, identify a low-dimensional subspace from data and project the governing equations onto it to obtain reduced dynamics, as in classical Proper Orthogonal Decomposition (POD)–Galerkin methods \citep{rempfer_pod_galerkin_2000}. When only measurements are available, data-driven identification methods are often used \citep{Brunton_ML_2020, Raissi_sysid_2018}. Traditional parametric input–output models, such as autoregressive models \citep{box_time_series_2015,Billings_narmax_2013}, or output-error formulations \citep{Ljung_oe_methods_2003}, remain popular due to their interpretability and computational efficiency. Gaussian process regression has also been employed for input-output maps \citep{Seeger_gp_204} with uncertainty quantification. Recent advances have introduced sparsity-promoting approaches such as the System Identification of Nonlinear Dynamics \citep[SINDy,][]{Brunton_sindy_2016, Brunton_sindy_control_2016} and neural-network-based architectures \citep{Chiuso_ml_sysid_2019, Schussler_rnn_sysid_2019}, which enable the identification of complex nonlinear behaviours directly from data. 
In parallel, Operator Inference \citep[OpInf,][]{Kramer_OpInf_2024, Peherstorfer_opinf_2016} has extended this idea in a data-driven, non-intrusive manner by identifying reduced operators directly from simulation or experimental data. Operator Inference has demonstrated success in learning reduced-order models even for complex partial differential equations \citep{Qian_OpInfPDE_2022} and experimental test cases \citep{Kramer_OpInf_2024}. More recently, autoencoders have also been used to map the system to a nonlinear latent space for learning dynamics \citep{Agostini_AE_predictions_2020, Zhang_autoencoder_sindy_2025}. While they can capture a strongly nonlinear behaviour, the nonlinear mapping is often complex and less interpretable than a linear projection. ROM techniques have been increasingly integrated into MPC frameworks to achieve real-time feasibility. Examples include MPC formulations based on DMD \citep{Lu_mpc_dmd_2021}, Spectral Submanifolds \citep[SSM,][]{Alora_ssm_mpc_2023}, and POD-based models \citep{Hovland_mpc_pod_2006, Karg_mhe_mpc_2021}.

%%%%%% The challenges in implementing MPC: the need for the state for prediction and action planning
However, even with reduced models, MPC of high-dimensional and nonlinear systems remains limited by model inaccuracies, truncation errors, and measurement sparsity. In most practical implementations, only sparse and noisy measurements are available, necessitating an estimation mechanism to reconstruct the full or latent system state. This leads to output-feedback MPC formulations \citep{Findeisen_output_feedback_mpc_2003}, in which control decisions are based on estimated rather than directly observed states. Common estimation strategies include the Kalman Filter and its nonlinear or ensemble-based extensions, such as Extended \citep{Ribeiro_EKF_2004}, Ensemble \citep{Evensen_EnKF_2003}, and Unscented (UKF) Kalman Filters \citep{Wan_UKF_2000}, or moving horizon estimation strategies \citep{Tenny_mpc_mhe_2002}.
An overview of output-feedback MPC is provided in \cite{Findeisen_output_feedback_mpc_2003}, while examples of applications using a Kalman Filter or moving horizon estimation are included in \cite{Lee_ekf_mpc_1994, Janiszewski_mpc_ukf_motor_2024, Tenny_mpc_mhe_2002}. 

Building upon these foundations, this work proposes an integrated framework that addresses the aforementioned challenges to provide a scalable framework for MPC of high-dimensional systems under partial observability. The key enablers of the framework are: (i) OpInf to learn the governing dynamics in a data-driven, non-intrusive manner directly in a latent space spanned by orthogonal POD modes; (ii) a state estimation performed with a UKF, which corrects the open-loop latent-state predictions of the OpInf model using sparse and noisy measurements, and (iii) MPC formulated directly in the latent space, substantially reducing computational cost while retaining nonlinear predictive capability. The framework does not require observing the full state at any moment once the model operators are inferred.
This methodology is demonstrated on the one- \citep{Papageorgiou_chaos_ks1d_1991} and two-\citep{Kalogirou_2DKS_2015} dimensional Kuramoto–Sivashinsky equations, canonical benchmarks for chaotic and spatially extended systems.
The proposed approach is general and can be extended to multi-dimensional flows or combined with alternative low-dimensional representations, including neural-network-based autoencoders.

The remainder of this paper is organised as follows.
Section \ref{sec:methodology} introduces the proposed methodology, detailing the implementation of the MPC formulation, the state estimation procedure based on the UKF, and the data-driven system identification using OpInf.
The results obtained for the one-dimensional Kuramoto–Sivashinsky equation are discussed in \cref{sec:results_1DKS}, while the extension to the two-dimensional case is presented in \cref{sec:results_2DKS}.
Finally, the main conclusions and perspectives of this work are summarised in \cref{sec:conclusions}.\\

\section{Model predictive control from partial measurements}
\label{sec:methodology}

%%%%%%%%%%% Framework descritption %%%%%%%%%%%

% Intro ...
This section introduces the proposed control framework, detailing the modelling strategy based on OpInf, the formulation of MPC, and the application of the UKF for state estimation from sparse and noisy measurements.

% The dynamical system ...
The process to be controlled is modelled as an input–output dynamical system characterised by the state vector $\boldsymbol{x}_k \in \mathbb{R}^{n_x}$ sampled at discrete time instants $t_k$, where $n_x$ denotes the dimension of the state space resulting from a spatial discretisation. The system evolves under the influence of a control input $\boldsymbol{u}_k \in \mathbb{R}^{n_u}$, with $n_u$ representing the dimension of the control space. The discrete-time dynamics are expressed as
\begin{equation}
\boldsymbol{x}_{k+1} = F(\boldsymbol{x}_k, \boldsymbol{u}_k),
\label{eq:system_dynamics}
\end{equation}
where $F: \mathbb{R}^{n_x} \times \mathbb{R}^{n_u} \to \mathbb{R}^{n_x}$ represents the nonlinear state-transition function, and $\boldsymbol{x}_0$ denotes the initial state. The discrete-time formulation is obtained via uniform sampling of the continuous-time system with a fixed sampling interval $\Delta t > 0$, such that $t_k = k \Delta t$. Throughout this work, $\boldsymbol{z}_k = \boldsymbol{z}(t_k)$  denotes a generic time-dependent variable $\boldsymbol{z}$ evaluated at $t_k$.

% What is control? ...
Formally, the control problem consists of determining the control input $\boldsymbol{u}_k$ at each discrete time instant $t_k$ such that a prescribed performance criterion is optimised subject to system dynamics and problem-dependent constraints. The objective is formulated through a cost function that encodes the desired control goals and which is optimised during the control process. In this work, the focus is on a class of problems where the objective is either to stabilise the system around a fixed reference point or to track a desired trajectory over time. Specifically, at each discrete time instant $t_k,~ k = 1, 2, \ldots$, the controller aims to steer the system state $\boldsymbol{x}_k$ towards a reference $\boldsymbol{x}^*_k$, which may be constant (\emph{i.e.}, $\boldsymbol{x}^*_k = \boldsymbol{x}^*,~\forall k$) or time-dependent. The cost function penalises deviations from the target state, while possibly also accounting for control effort or other constraints depending on the application.

\begin{figure}
\centering
\includegraphics[]{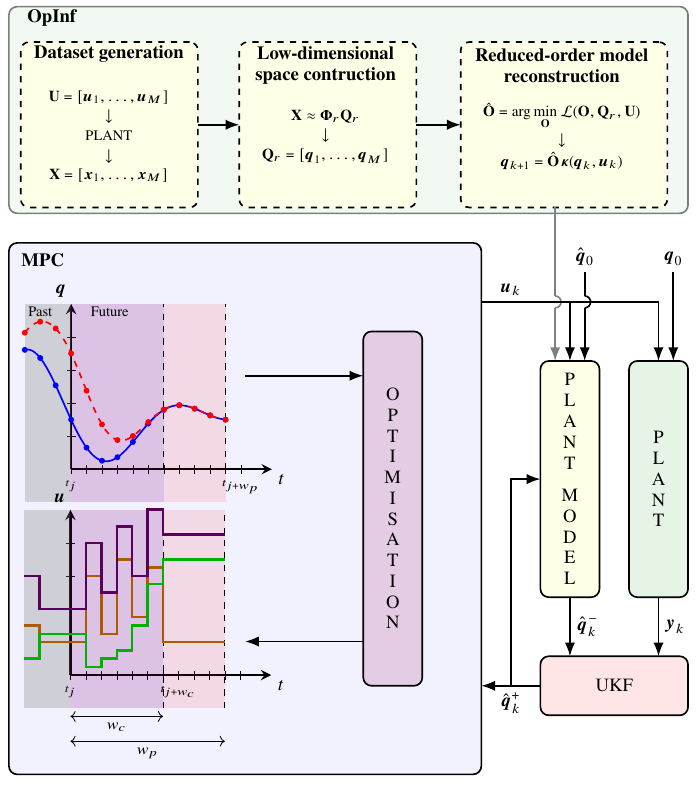}
\caption{\justifying Schematic of the control algorithm.
\textbf{Top:} Data-driven model identification. An offline dataset of trajectories with $M$ snapshots is collected with open-loop control, and a low-dimensional latent space is constructed from dominant Proper Orthogonal Decomposition modes. The system dynamics are learned directly in this latent space via Operator Inference in a non-intrusive manner and including nonlinear dependencies. 
\textbf{Bottom:} Control procedure. An Unscented Kalman Filter estimates the latent coordinates using partial and noisy state measurements to correct the model predictions. The posterior Kalman estimate is used to initialise the predictive model at each control step. Within Model Predictive Control, the surrogate model generates open-loop latent-state predictions (solid blue line; dots indicate discrete time samples) over a prediction horizon $w_p$ (red shaded area). These predictions are used to optimise the control sequence over the control horizon $w_c$ (purple shaded area). The dashed red line indicates the target latent trajectory.}
\label{fig:control-schematic}
\end{figure}

% We use MPC to do this...
To address this problem, MPC is employed as a model-based approach. This control technique relies on an explicit representation of the system dynamics to predict its future evolution under a sequence of control actions. These open-loop predictions are used to directly evaluate a cost function, which is then optimised online. The result is an optimal sequence of control actions defined over the prediction horizon. Only the first action in this sequence is applied to the system, while the remaining are discarded. Once the system state is updated, the optimisation is repeated over a shifted horizon. A detailed description of the MPC algorithm, including its formulation and implementation, is presented in Subsection~\ref{subsec:MPC}, while the overall MPC procedure is illustrated schematically on the bottom-left side of \cref{fig:control-schematic}.

%% Figure 1: schematic of the control algorithm

% Which is the model we use for MPC?...
For accurate cost-function evaluation via model rollouts within the MPC loop, it is essential to have a reliable prediction model of the system dynamics. The model must satisfy two main requirements: (i) it should accurately capture the temporal evolution of the system over the prediction horizon, and (ii) it must be computationally efficient to allow real-time implementation, particularly in high-dimensional settings. In this work, the surrogate model is obtained via OpInf \citep{Kramer_OpInf_2024}. The full-order states are projected onto a low-dimensional subspace defined by the leading POD modes, yielding a latent coordinate $\boldsymbol{q}_k \in \mathbb{R}^{r}$ at time step $t_k$, where $r\ll n_x$ denotes the reduced dimension. The system dynamics are then learned directly in the latent space by imposing a polynomial form (see the top of \cref{fig:control-schematic}). With dynamics defined in the latent space, and owing to the orthonormality of the POD basis and linear mapping, the control problem can be formulated directly in this reduced space, greatly decreasing the computational cost of the MPC implementation.

% The state is generally not available in the closed loop...
In practical scenarios, the controller has access to partial and noisy measurements of the state. These are modelled as
\begin{equation}
    \boldsymbol{y}_k = C(\boldsymbol{x}_k) + \boldsymbol\nu^y_k \quad \in \mathbb{R}^{n_y},
    \label{eq:measure}
\end{equation} 
where $n_y$ is the dimensionality of the measurement vector, $C: \mathbb{R}^{n_x} \to \mathbb{R}^{n_y}$ is a known operator that maps the full state to observable quantities, and $\boldsymbol\nu^y_k$ denotes the realisation of the measurement noise at time $t_k$.

% We then need a state estimator to close the loop
To close the feedback loop, a state estimator is then needed to infer the current latent state from the available measurements. In the proposed framework, a UKF is used for this purpose. In particular, this is used in the MPC loop to perform the predictions for the estimation of the cost function. The latent coordinate vector $\boldsymbol{q}^-_k$ coming from the prediction of the Opinf-based model is corrected directly from the sparse and possibly noisy measurements $\boldsymbol{y}_k$ available during control to obtain a posterior Kalman estimate $\boldsymbol{q}^+_k$. This approach accounts for both model uncertainty and measurement noise, enabling accurate prediction of the latent state for MPC rollouts. Unfortunately, unlike the linear case, a separation principle does not generally hold for nonlinear systems, so independent designs of controllers and observers do not guarantee closed-loop stability \citep{Findeisen_output_feedback_mpc_2003}. However, MPC is inherently robust to small perturbations in the state estimate, and as long as the state estimator performs well, MPC can still be applied effectively in practical scenarios; otherwise, robust MPC implementations \citep{Mayne_robust_output_feedback_mpc_2006} are needed.

%%%%%%%%%%% OPINF %%%%%%%%%%%
\subsection{Data-driven plant model estimation}
\label{subsec:Opinf}

Among the methodologies useful to derive a surrogate plant model, OpInf provides a non-intrusive data-driven approach. In this framework, the governing equations of the system are neither assumed to be known nor used directly. Instead, the model is inferred purely in a data-driven manner, from snapshots of state trajectories collected from simulations or experiments.

The OpInf procedure can be structured into three main steps: (i) dataset generation, (ii) low-dimensional space construction, and (iii) reduced-order model construction. The resulting ROM can then be integrated within the MPC framework to predict future system states and to optimise control inputs for efficient control.

\subsubsection{Dataset generation}
In this phase, a dataset of states and inputs at different time instants is obtained. Specifically, a set of open-loop control actions is first selected, and the system is then either simulated using the true system's dynamics in \cref{eq:system_dynamics} or measured experimentally to collect the resulting state trajectories. The dataset, composed of $M$ snapshots collected from one or more open-loop trajectories, is arranged into snapshot matrices as follows:
\begin{equation}
   \mathbf{U} = \left[ \boldsymbol{u}_1, \ldots,  \boldsymbol{u}_{M} \right] \in \mathbb{R}^{n_u \times M}, \quad
    \mathbf{X} = \left[ \boldsymbol{x}_1, \ldots, \boldsymbol{x}_{M} \right] \in \mathbb{R}^{n_x \times M}
    \label{eq:snapshots_matrix}
\end{equation}
where the matrix $\mathbf{U}$ contains the selected open-loop control inputs and $\mathbf{X}$ the states sampled.

\subsubsection{Low-dimensional space construction}
A reduced-order basis is extracted from the high-dimensional system states using POD. This technique relies on the singular value decomposition of the snapshot matrix $\mathbf{X}$:
\begin{equation}
\mathbf{X} = \mathbf{\Phi} \boldsymbol{\Sigma} \mathbf{\Psi}^\top,
\label{eq:POD}
\end{equation}
where the superscript $\top$ denotes the transposition operation, $\mathbf{\Phi} = [\boldsymbol{}{\phi}_1, \ldots, \boldsymbol{}{\phi}_{n_x}] \in \mathbb{R}^{n_x \times n_x}$ contains the left singular vectors, representing the spatial modes; $\boldsymbol{\Sigma} = \mathrm{diag}(\sigma_1, \ldots, \sigma_M) \in \mathbb{R}^{n_x \times M}$ is the diagonal matrix of non-negative singular values $\sigma_i, ~i = 1, \ldots,M$, ordered in descending magnitude; and $\mathbf{\Psi} = [\boldsymbol{\psi}_1, \ldots, \boldsymbol{\psi}_M] \in \mathbb{R}^{M \times M}$ contains the right singular vectors, encoding the temporal structure. The columns of $\mathbf{\Phi}$ and $\mathbf{\Psi}$ form orthonormal bases, satisfying $\mathbf{\Phi}^\top \mathbf{\Phi} = \mathbf{I}_{n_x}$ and $\mathbf{\Psi}^\top \mathbf{\Psi} = \mathbf{I}_M$, where $\mathbf{I}_{d}$ is the identity matrix of dimension $d$.

By retaining only the first $r \ll \mathrm{min}(n_x,M)$ dominant singular values and vectors, a rank-$r$ approximation of the snapshot matrix is obtained:
\begin{equation}
\mathbf{X} \approx \mathbf{\Phi}_r \boldsymbol{\Sigma}_r \mathbf{\Psi}_r^\top = \mathbf{\Phi}_r \mathbf{Q}_r,
\label{eq:POD_truncated}
\end{equation}
where $\mathbf{\Phi}_r \in \mathbb{R}^{n_x \times r}$, $\boldsymbol{\Sigma}_r \in \mathbb{R}^{r \times r}$ and $\mathbf{\Psi}_r \in \mathbb{R}^{M \times r}$ are the matrices containing the first $r$ left singular vector, singular values and right singular vectors, respectively. The matrix $\mathbf{Q}_r = \boldsymbol{\Sigma}_r \mathbf{\Psi}_r^\top \in \mathbb{R}^{r\times M}$ represents the modal amplitudes, obtained by weighting each right singular vector by its corresponding singular value. A common approach for selecting the dimension of the latent space $r$ is to observe the decay of the singular values. Specifically, the retained variance associated with the first $r$ modes is computed as the normalised cumulative sum of the squared singular values $\gamma_r = \sum_{i=1}^{r} \sigma_i^2 / \sum_{i=1}^{\mathrm{min}(n_x,M)} \sigma_i^2$. The value of $r$ is typically chosen such that it exceeds a predefined threshold, ensuring that most of the system's variance is captured \citep{brindise_podtrunc_2017, Kramer_OpInf_2024}.

Once $r$ has been selected, the basis matrix $\mathbf{\Phi}_r$, spanning an $r$-dimensional subspace, defines a low-dimensional coordinate system used to construct the ROM. Each high-dimensional state vector $\boldsymbol{x}_k$ is projected onto the reduced basis as:
\begin{equation}
\boldsymbol{q}_k = \mathbf{\Phi}_r^\top \boldsymbol{x}_k.
\label{eq:decoding}
\end{equation}
where $\boldsymbol{q}_k \in \mathbb{R}^{r}$ corresponds to the low-dimensional representation of $\boldsymbol{x}_k$ and one has used the fact that $\mathbf{Q}_r$ defines an orthonormal basis.
A low-order reconstruction of the vector $\boldsymbol{x}_k$, denoted as $\tilde{\boldsymbol{x}}_k$, can then be obtained as
\begin{equation}
\boldsymbol{x}_k \approx \mathbf{\Phi}_r \boldsymbol{q}_k = \tilde{\boldsymbol{x}}_k \in \mathbb{R}^{n_x}.
\label{eq:encoding}
\end{equation}
The matrix $\mathbf{Q}_r$ of modal amplitudes is the collection of all the reduced state representations $\mathbf{Q}_r = \left[\boldsymbol{q}_1,\ldots,\boldsymbol{q}_M\right]$ in the training dataset.
%In the following, $\tilde{\boldsymbol{x}}_k$ will denote the low-order reconstruction of a known state ${\boldsymbol{x}}_k$. 

\subsubsection{Reduced-order model construction}
Once the reduced basis $\mathbf{\Phi}_r$ has been identified, the next step consists of learning a dynamical model that evolves the system in the reduced-order space. The discrete-time reduced dynamics to be learned have the following form:
\begin{equation}
	\boldsymbol{q}_{k+1} = F_q(\boldsymbol{q}_k, \boldsymbol{u}_k),
\end{equation}
where $F_q: \mathbb{R}^{r} \times \mathbb{R}^{n_u} \to \mathbb{R}^{r}$ describes the temporal evolution of the latent coordinates and must be inferred from data. Operator Inference provides a data-driven framework to identify $F_q$ by fitting a polynomial model to snapshot data. In particular, the inference is performed in a polynomial feature space:
\begin{equation}
	\boldsymbol{q}_{k+1} = \boldsymbol{c} + \mathbf{A}\boldsymbol{q}_k + \mathbf{H}(\boldsymbol{q}_k \circledast \boldsymbol{q}_k) + \mathbf{G}(\boldsymbol{q}_k \circledast \boldsymbol{q}_k \circledast \boldsymbol{q}_k) + \mathbf{B}\boldsymbol{u}_k + \mathbf{N}(\boldsymbol{q}_k \circledast \boldsymbol{u}_k) + \ldots,
    \label{eq:model_deps}
\end{equation}
where $\boldsymbol{c}$ is a constant offset, $\mathbf{A}$ is the linear operator, $\mathbf{H}$ and $\mathbf{G}$ encode quadratic and cubic nonlinearities, respectively, $\mathbf{B}$ and $\mathbf{N}$ model the input and state-input interaction terms, and other terms can be included. The symbol $\circledast$ denotes the symmetrised Kronecker product used to construct higher-order tensor features. For  $\boldsymbol{q}\in \mathbb{R}^r$, the degree-$n$ polynomial term, $\boldsymbol{q} \circledast \overset{n}{\cdots} \circledast \boldsymbol{q}$, is a vector of dimension $\binom{r+n-1}{n}$.

To express this model compactly, a feature operator $\boldsymbol{\rho}: \mathbb{R}^r \times \mathbb{R}^{n_u} \to \mathbb{R}^m$ is defined, where $m$ represents the total number of polynomial features constructed from the latent variables and control inputs up to the chosen polynomial order. This operator maps the latent state and control input to the corresponding polynomial basis:
\begin{equation}
	\boldsymbol{\rho}(\boldsymbol{q}_k, \boldsymbol{u}_k) := 
	\begin{bmatrix}
		1 \\
		\boldsymbol{q}_k \\
		\boldsymbol{q}_k \circledast \boldsymbol{q}_k \\
		\boldsymbol{q}_k \circledast \boldsymbol{q}_k \circledast \boldsymbol{q}_k \\
		\boldsymbol{u}_k \\
		\boldsymbol{q}_k \circledast \boldsymbol{u}_k \\
		\vdots
	\end{bmatrix} \in \mathbb{R}^m.
\end{equation}
The reduced-order dynamics can then be written as
\begin{equation}
	\boldsymbol{q}_{k+1} = \mathbf{O} \, \boldsymbol{\rho}(\boldsymbol{q}_k, \boldsymbol{u}_k),
\end{equation}
where $\mathbf{O} \in \mathbb{R}^{r \times m}$ is the matrix, to be identified, that collects all the operators from \cref{eq:model_deps} and is structured as
\begin{equation}
	\mathbf{O} = \begin{bmatrix}
		\boldsymbol{c} & \mathbf{A} & \mathbf{H} & \mathbf{G} & \mathbf{B} & \mathbf{N} & \ldots
	\end{bmatrix}.
    \label{eq:op_matrix}
\end{equation}
To infer $\mathbf{O}$ from data, the feature operator $\boldsymbol{\rho}$ is evaluated offline on snapshots $(\boldsymbol{q}_k, \boldsymbol{u}_k)$ to form the design matrix:
\begin{equation}
	\mathbf{D} = \begin{bmatrix}
		\boldsymbol{\rho}(\boldsymbol{q}_1, \boldsymbol{u}_1)^\top \\
		\vdots \\
		\boldsymbol{\rho}(\boldsymbol{q}_{M-1}, \boldsymbol{u}_{M-1})^\top
	\end{bmatrix} \in \mathbb{R}^{(M-1) \times m},
\end{equation}
and the corresponding target matrix:
\begin{equation}
	\mathbf{Z} = \begin{bmatrix}
		\boldsymbol{q}_{2}^\top \\
		\vdots \\
		\boldsymbol{q}_{M}^\top
	\end{bmatrix} \in \mathbb{R}^{(M-1) \times r}.
\end{equation}
The OpInf problem is then described by the following regularised least-squares optimisation:
\begin{equation}
	\hat{\mathbf{O}} = \arg\min_{\mathbf{O}} \| \mathbf{D} \mathbf{O}^\top - \mathbf{Z} \|_F^2 + \lambda_o \| \mathbf{O} \|_F^2,
	\label{eq:OpInf_optimization}
\end{equation}
where $\lambda_o \ge 0$ is a regularization parameter and $\|\cdot\|_F$ denotes the Frobenius norm. The first term in \cref{eq:OpInf_optimization} promotes model accuracy with respect to the training data, while the second discourages overfitting by penalising large operator coefficients. In particular, the regularisation term acts as a bias in the optimisation problem that guides learning when data are scarce or noisy. This promotes meaningful models with reasonable predictive capabilities that can generalise well to unseen inputs and initial conditions.

Once the optimal operator $\hat{\mathbf{O}}$ has been identified, the OpInf model is defined by:
\begin{equation}
	\hat{\boldsymbol{q}}_{k+1} = \hat{\mathbf{O}} \, \boldsymbol{\rho}(\hat{\boldsymbol{q}}_k, \boldsymbol{u}_k),
	\label{eq:OpInf_model}
\end{equation}
where $\hat{\boldsymbol{q}}_k$ denotes the estimated latent states from the reduced-order model, and the corresponding full-state reconstruction is obtained as $\hat{\boldsymbol{x}}_k = \mathbf{\Phi}_r \hat{\boldsymbol{q}}_k$.

\subsection{Model Predictive Control}
\label{subsec:MPC}
In this work, MPC is formulated directly in terms of the latent variables identified via OpInf. The following formulation assumes access to estimated latent states at each control step and leverages the reduced-order dynamics for control purposes.

In particular, starting from time instants $t_k$, given an estimate of the current latent coordinate $\hat{\boldsymbol{q}}_k$, the model predicts future latent coordinates $\hat{\boldsymbol{q}}_{k+j|k}$ under a sequence of control actions over a prediction horizon $t_{k+j}, j = 1,\ldots,w_p$. The notation $\hat{\boldsymbol{q}}_{k+j|k}$ indicates a prediction of the latent coordinate at time step $t_{k+j}$ obtained advancing the learned predictive model in \cref{eq:OpInf_model} from the latent state estimate $\hat{\boldsymbol{q}}_k$ , such that $\hat{\boldsymbol{q}}_{k|k} := \hat{\boldsymbol{q}}_k$.

The optimal input sequence $\{\boldsymbol{u}_{k+j}^{\mathrm{opt}}\}_{j=0}^{w_c-1}$ is then obtained by minimising in a control window $t_{k+j}, j = 0,\ldots,w_c-1$, the cost function
\begin{equation}   
\begin{split}
J(\hat{\boldsymbol{u}}_{k}, \ldots, \hat{\boldsymbol{u}}_{k+w_c-1}, \hat{\boldsymbol{q}}_k) = & 
\sum_{j=0}^{w_p-1} \| \hat{\boldsymbol{q}}_{k+j|k} - \boldsymbol{q}^*_{k+j} \|^2_{\mathbf{R}_q}  + V_f(\hat{\boldsymbol{q}}_{k+w_p|k})\\
+ & \sum_{j=0}^{w_c-1} \left( 
\| \hat{\boldsymbol{u}}_{k+j} \|^2_{\mathbf{R}_u} 
+ \| \Delta \hat{\boldsymbol{u}}_{k+j} \|^2_{\mathbf{R}_{\Delta u}} 
\right),
\end{split}
\label{eq:cost_function_mpc}
\end{equation}
with respect to the actions $\hat{\boldsymbol{u}}_{k}, \ldots, \hat{\boldsymbol{u}}_{k+w_c-1}$ used to build it via rollout. Here, $\| \boldsymbol{d} \|_{\mathbf{K}}^2 = \boldsymbol{d}^\top \mathbf{K} \boldsymbol{d}$ is the weighted norm of a generic vector $\boldsymbol{d}$ with respect to a symmetric and positive definite matrix $\mathbf{K}$. The matrices $\mathbf{R}_q$, $\mathbf{R}_u$, and $\mathbf{R}_{\Delta u}$ are weighting matrices assigning relative importance to each term in the cost function, with $\mathbf{R}_q$ typically positive semidefinite and $\mathbf{R}_u$, $\mathbf{R}_{\Delta u}$ positive definite. The function $V_f$ represents a terminal cost accounting for the desired performance at the end of the prediction horizon. This term captures contributions to the overall cost not explicitly considered in the accumulated running cost and renders MPC a local, model-based approximation of the optimal Bellman policy \citep{Bertsekas_mpc_rl_2024}. The term $\Delta \hat{\boldsymbol{u}}_{k+j|k} = \hat{\boldsymbol{u}}_{k+j|k} - \hat{\boldsymbol{u}}_{k+j-1|k}$ is the input variability in time. In the cost function, $\boldsymbol{q}^*_k$ denotes the low dimensional representation of the target state $\boldsymbol{x}^*$ at time $t_k$, obtained via projection $\boldsymbol{q}^*_k = \mathbf{\Phi}_r^\top\boldsymbol{x}^*_k$, as described in \cref{eq:decoding}. In general, $w_c \leq w_p$; when $w_c < w_p$, the control input is maintained constant beyond the control horizon.

After the optimisation, only the first control input in the sequence, $\boldsymbol{u}_{k} = \boldsymbol{u}^{\rm opt}_{k}$, is applied to the system, and the process identically repeats at the next time step and with the new latent state estimate available. MPC allows a straightforward introduction of state or input constraints within the optimisation in \cref{eq:cost_function_mpc}. For instance, indicating with superscript $(i)$ the $i$th component of a vector, constraints on the input are considered:
\begin{equation}
\hat{\boldsymbol{u}}^{(i)}_k \in [\boldsymbol{u}_{\min}, \boldsymbol{u}_{\max}], 
\quad 
\Delta\hat{\boldsymbol{u}}^{(i)}_k \in [\Delta \boldsymbol{u}_{\min}, \Delta \boldsymbol{u}_{\max}]
\quad 
\forall\,\, i,k.    
\end{equation}
% Here, the superscript $(i)$ denotes the $i$th component of a vector. 
The MPC loop is then summarised in Alg.~\ref{alg:mpc_optimization}.

\begin{algorithm}
\caption{MPC Optimization Problem}
\label{alg:mpc_optimization}
\begin{algorithmic}[1]
\State Collect the latent state estimate $\hat{\boldsymbol{q}}_k$ at time step $t_k$
\State Solve the optimisation:
\[
{\boldsymbol{u}}^{\mathrm{opt}}_{k}, \dots, {\boldsymbol{u}}^{\mathrm{opt}}_{k+w_c-1}
= \min_{\{\hat{\boldsymbol{u}}_{k}, \dots, \hat{\boldsymbol{u}}_{k+w_c-1}\}} 
J(\hat{\boldsymbol{u}}_k, \dots, \hat{\boldsymbol{u}}_{k+w_c-1}, \hat{\boldsymbol{q}}_k)
\]
Subject to constraints:
\[
\begin{array}{llll}
& \hat{\boldsymbol{u}}^{(i)}_{k+j} \in [\boldsymbol{u}_{\min}^{(i)}, \boldsymbol{u}_{\max}^{(i)}], & j = 0,\dots,w_c-1, & i = 1,\dots,m \\
& \Delta \hat{\boldsymbol{u}}^{(i)}_{k+j} 
\in [\Delta \boldsymbol{u}_{\min}^{(i)}, \Delta \boldsymbol{u}_{\max}^{(i)}], & j = 0,\dots,w_c-1, & i = 1,\dots,m\\ 
& \hat{\boldsymbol{u}}_{k+j} = \hat{\boldsymbol{u}}_{k+w_c-1},  & j \ge w_c &\\
& \hat{\boldsymbol{q}}_{k|k} := \hat{\boldsymbol{q}}_k, \\
& \hat{\boldsymbol{q}}_{k+j+1|k} = \hat{F}_q(\hat{\boldsymbol{q}}_{k+j|k}, \hat{\boldsymbol{u}}_{k+j}), &j = 0,\dots,w_p-1&
\end{array}
\]
\State Apply $\boldsymbol{u}_k = \boldsymbol{u}^{\mathrm{opt}}_{k}$
\State Update the time step to  $t_{k+1}$ and repeat from step $1$
\end{algorithmic}
\end{algorithm}

For all cases considered in this work, the optimisation problem in \cref{eq:cost_function_mpc} is solved using CasADi \citep{Andersson_Casadi_2019}, a symbolic framework for algorithmic differentiation that enables the efficient implementation of numerical optimal control methods, including nonlinear MPC. In particular, the optimisation problems are solved using IPOPT \citep[Interior Point OPTimizer,][]{Wachter_IPOPT_2006}, a large-scale nonlinear solver based on a primal-dual interior-point method.

\subsection{Data assimilation with UKF}
\label{subsec:UKF}
Once the plant model has been identified and the MPC framework established, it is necessary to initialise the predictive model with an appropriate estimate of the system's latent state $\hat{\boldsymbol{q}}_k$, at each control time step. This is essential to evaluate the cost function in \cref{eq:cost_function_mpc} via plant model rollouts. In realistic scenarios, MPC typically operates under partial observability, where only a measure $\boldsymbol{y}_k$ of the system state is available at time $t_k$. Consequently, reconstructing the full latent state vector $\boldsymbol{q}_k$ from $\boldsymbol{y}_k$ represents a state estimation problem. Note that measurements need not be collected at every control step; the estimation can propagate the state in between the updates using the model dynamics. To deal with this estimation problem, an Unscented Kalman Filter \citep{Julier_UKF_1997,Wan_UKF_2000} is employed. The UKF serves as a data assimilation layer, providing an online estimation of the latent state by combining the predictive reduced order model from Opinf in \cref{eq:OpInf_model} with the available observations. At each control time, the UKF first propagates the prior estimate of the latent coordinate forward in time using the model dynamics and then corrects this prediction using the current observation $\boldsymbol{y}_k$. This process allows a robust initialisation of the MPC at each decision step.

Although the true underlying system is deterministic, the learned dynamical model is an approximation subject to modelling errors, due to limited training data, truncation of the reduced-order basis, unmodeled dynamics, or imperfect identification of nonlinear interactions. As a result, these prediction errors can be represented as a stochastic process noise term, leading to the modified latent dynamics
\begin{equation}
	\hat{\boldsymbol{q}}_{k+1} = \hat{F}_q(\hat{\boldsymbol{q}}_k, \boldsymbol{u}_k) + \boldsymbol{\nu}^q_k
\end{equation} 
where $\boldsymbol{\nu}^q_k$ is a realisation of the process noise at time $t_k$, representing the stochastic component of the model and accounting for unmodeled residuals.

The Kalman Filter addresses the state estimation problem by assuming the latent coordinate is a Gaussian random variable. The corrected estimate at time $t_k$ is computed as
\begin{equation}
	\hat{\boldsymbol{q}}_k^+ = \hat{\boldsymbol{q}}_k^- + \boldsymbol{K}_k \left( \boldsymbol{y}_k - \hat{\boldsymbol{y}}_k \right),
\end{equation}
where $\hat{\boldsymbol{q}}_k^-$ is the prior latent state prediction obtained by integrating the OpInf surrogate model in \cref{eq:OpInf_model} and $\hat{\boldsymbol{y}}_k$ is the predicted current observation derived through the measurement function defined in \cref{eq:measure}. The matrix $\boldsymbol{K}_k \in \mathbb{R}^{r \times n_y}$ is the Kalman gain and determines how much the prior estimate is adjusted in response to the observed innovation $(\boldsymbol{y}_k - \hat{\boldsymbol{y}}_k)$.

Following the Kalman Filter theory (see \cite{Welch_KF_1995}), the optimal computation of the Kalman gain is given by:
\begin{equation}
	\boldsymbol{K}_k = \mathbf{P}_{q y}\mathbf{P}_{y y}^{-1},
\end{equation}
where $\mathbf{P}_{q y} \in \mathbb{R}^{r \times n_y}$ and $\mathbf{P}_{y y} \in \mathbb{R}^{n_y \times n_y}$ denote the cross-covariance matrices between the latent state and predicted measurement,  and the covariance of the predicted measurement, respectively. Here, $\mathbf{P}_{y y}^{-1}$ denotes the inverse of the matrix $\mathbf{P}_{y y}$. In linear systems, these quantities can be computed analytically, yielding an exact recursive solution. However, for nonlinear systems, these covariances involve expectations over nonlinear transformations of random variables and cannot be obtained in closed form. The Extended Kalman Filter \citep[EKF,][]{Ribeiro_EKF_2004} addresses this by linearising the dynamics and observation models about the current estimate, using a first-order Taylor expansion to approximate the required covariances. This approach may lead to sub-optimal performance or even divergence of the filter, since the first-order linearization does not accurately capture the true nonlinear transformations of the mean and covariance, resulting in incorrect propagation of the state uncertainty \citep{Marafioti_nonlinearmpc_2009}.

To overcome these limitations, the UKF uses the unscented transformation (UT), which deterministically selects a set of points (called sigma points) that are propagated through the nonlinear model and measurement functions. These sample points completely capture the true mean and covariance of the latent state and, when propagated through the nonlinear system, capture the posterior mean and covariance accurately to the third order for any nonlinearity. Alternatively, the Ensemble Kalman Filter \citep[EnKF,][]{Evensen_EnKF_2003} employs a Monte Carlo approach that propagates an ensemble of system realisations to estimate the posterior statistics. The EnKF has been used in a similar control problem in reinforcement learning \citep{Ozan_DataAssimilationRL_2025}.

%%% UKF steps
The UKF estimation process comprises four main steps: (i) filter initialisation, (ii) sigma point generation, (iii) time update, and (iv) measurement update and correction.

\subsubsection{Filter initialisation}
To incorporate both process and measurement noise within the estimation framework, an augmented state vector is defined at each time step $t_k$:
\begin{equation}
	\boldsymbol{q}_{k}^a = \begin{bmatrix}
		\boldsymbol{q}_{k} \\
		\boldsymbol{\nu}^q_{k} \\
		\boldsymbol{\nu}^y_{k}
	\end{bmatrix} \in \mathbb{R}^{n_a},
\end{equation}
with $n_a = 2r + n_y$ the augmented dimension.
Similarly, the augmented state covariance matrix is built as
\begin{equation}
	\mathbf{P}_{k}^a = \begin{bmatrix}
		\mathbf{P}_{k}^+ & \mathbf{0} & \mathbf{0} \\
		\mathbf{0} & \mathbf{Q} & \mathbf{0} \\
		\mathbf{0} & \mathbf{0} & \mathbf{R}
	\end{bmatrix} \in \mathbb{R}^{n_a \times n_a},
\end{equation}
where $\mathbf{P}_{k}^+ \in \mathbb{R}^{r \times r}$ represents the posterior covariance matrix of the latent state, while $\mathbf{Q} \in \mathbb{R}^{r \times r}$ and $\mathbf{R} \in \mathbb{R}^{n_y \times n_y}$ denote the known covariance matrices of the process and measurement noise, respectively.

At the initial time $t_0$, the UKF is started with an initial estimate of the latent coordinate posterior mean, $\hat{\boldsymbol{q}}_{0}^{+}$ and covariance matrix $\mathbf{P}_{0}^{+}$:
\begin{align}
{\hat{\boldsymbol{q}}}_{0}^{+} &= \mathbb{E}[\boldsymbol{q}_0]\\
\mathbf{P}_{0}^{+} &= \mathbb{E}[(\boldsymbol{q}_0 - {\hat{\boldsymbol{q}}}_{0}^{+})(\boldsymbol{q}_0 - {\hat{\boldsymbol{q}}}_{0}^{+})^\top],
\end{align}
where $\mathbb{E}[\cdot]$ denotes the expectation operator. Since these latent quantities are not directly observable, they are obtained indirectly. First, the full initial state $\hat{\boldsymbol{x}}_0^{+}$ is estimated from $\boldsymbol{y}_0$ using Gaussian Process Regression \citep[GPR,][]{Schulz_gpr_2018}. This estimate is then projected onto the reduced basis to yield the latent initial state $\hat{\boldsymbol{q}}_0^{+} = \mathbf{\Phi}_r^\top \hat{\boldsymbol{x}}_0^{+}$. The corresponding latent covariance, $\mathbf{P}_0^{+}$, is derived from the uncertainty provided by the GPR model, ensuring a consistent initialisation for recursive filtering.  

Then, the augmented state is initialised as 
\begin{equation}
	\hat{\boldsymbol{q}}_{0}^{a} = [\begin{matrix} (\hat{\boldsymbol{q}}_{0}^{+})^\top & \boldsymbol{0}_r^\top & \boldsymbol{0}_{n_y}^\top \end{matrix}]^\top,
\end{equation}
where $\boldsymbol{0}_{d}$ denotes a zeros vector of dimension $d$. Similarly, the augmented state covariance matrix is
\begin{equation}
	\mathbf{P}_{0}^{a} = \begin{bmatrix} \mathbf{P}_{0}^{+} & 0 & 0 \\ 0 & \mathbf{Q} & 0 \\ 0 & 0 & \mathbf{R} \end{bmatrix}.
\end{equation}

\subsubsection{Sigma points generation}
A set of $2 n_a + 1$ sigma points is generated to approximate the distribution of the augmented state:
\begin{equation}
	\begin{aligned}
		\boldsymbol{\mathcal{Q}}_{k-1}^{a,(0)} &= \hat{\boldsymbol{q}}_{k-1}^{a}, \\
		\boldsymbol{\mathcal{Q}}_{k-1}^{a,(i)} &= \hat{\boldsymbol{q}}_{k-1}^{a} + \sqrt{n_a+\lambda} \: \mathbf{S}_{(\cdot, i)}, \quad \forall~ i=1, \ldots, n_a, \\
		\boldsymbol{\mathcal{Q}}_{k-1}^{a,(i)} &= \hat{\boldsymbol{q}}_{k-1}^{a} - \sqrt{n_a+\lambda} \: \mathbf{S}_{(\cdot, i)}, \quad \forall~ i= n_a+1, \ldots, 2n_a,
	\end{aligned}
\end{equation}
where $\mathbf{S}$ is the Cholesky factor of $\mathbf{P}_{k-1}^{a}$ and $\lambda = \alpha^2(n_a + \kappa) - n_a$, with $\alpha$ generally a small value $0<\alpha\leq 1$ and $\kappa \geq 0$ tuning parameters controlling the spread of the sigma points.

For simplicity, the sigma points are partitioned into these components:
\begin{equation}
	\boldsymbol{\mathcal{Q}}_{k-1}^{a,(i)} = \begin{bmatrix} \boldsymbol{\mathcal{Q}}_{k-1}^{q,(i)}\\
	\boldsymbol{\mathcal{Q}}_{k-1}^{\nu^q,(i)}\\
	\boldsymbol{\mathcal{Q}}_{k-1}^{\nu^y,(i)}\\
 \end{bmatrix}.
\end{equation}
where the superscripts $q$, $\nu^q$, $\nu^y$ refer to a partition conformal to the dimensions of the latent state, process noise, and measurement noise, respectively.

\subsubsection{Time update}
Each sigma point is propagated through the dynamic model:
\begin{equation}
    \boldsymbol{\mathcal{Q}}_{k}^{q-, (i)} = \hat{F}_q(\boldsymbol{\mathcal{Q}}_{k-1}^{q, (i)}, \boldsymbol{u}_{k-1}) +  \boldsymbol{\mathcal{Q}}_{k-1}^{\nu^q,(i)}, \quad \forall i = 0, \ldots, 2n_a + 1,
\end{equation}
where $\boldsymbol{\mathcal{Q}}_{k}^{q-, (i)}$ denotes the one-step prediction of the $i$th sigma point $\boldsymbol{\mathcal{Q}}_{k-1}^{a,(i)}$ using the surrogate model of the dynamics in \cref{eq:OpInf_model}.
The predicted prior mean $\hat{\boldsymbol{q}}_{k}^{-}$ and covariance $\mathbf{P}_{k}^{-}$ of the latent state at time $k$ are approximated by weighted sums of the propagated sigma points:
\begin{align}
	\hat{\boldsymbol{q}}_{k}^{-} &= \sum_{i=0}^{2n_a} W_i^{m} \boldsymbol{\mathcal{Q}}_{k}^{q-, (i)}\\
	\mathbf{P}_{k}^{-} &= \sum_{i=0}^{2n_a} W_i^{c} \left[\boldsymbol{\mathcal{Q}}_{k}^{q-, (i)} -  \hat{\boldsymbol{q}}_{k}^{-}\right]\left[\boldsymbol{\mathcal{Q}}_{k}^{q-, (i)} -  \hat{\boldsymbol{q}}_{k}^{-}\right]^\top
\end{align}
where the weighting coefficients are defined as
\begin{align}
	W_0^{m} &= \frac{\lambda}{n_a+\lambda}\\
	W_0^{c} &= \frac{\lambda}{n_a+\lambda} + (1-\alpha^2+\beta)\\
	W_i^{m} &= W_i^{c} = \frac{1}{2(n_a+\lambda)} \quad \forall i=1, \ldots, 2n_a
\end{align}
being $\beta$ a parameter incorporating prior knowledge of the distribution of the state. For Gaussian distributions, $\beta = 2$ is optimal.

\subsubsection{Measurement update and correction}
Each sigma point is transformed through the measurement function included in \cref{eq:measure} to the observation space:
\begin{equation}
	\boldsymbol{y}_{k}^{-, (i)} = C(\mathbf{\Phi}_r\boldsymbol{\mathcal{Q}}_{k}^{q-, (i)}) + \boldsymbol{\mathcal{Q}}_{k-1}^{\nu^y, (i)}
\end{equation}
The predicted measurement mean, covariance and cross-covariance are then computed as:	
\begin{align}
	\hat{\boldsymbol{y}}_{k}^{-} &= \sum_{i=0}^{2n_a} W_i^{m} \boldsymbol{y}_{k}^{-, (i)}\\
	\mathbf{P}_{yy,k} &= \sum_{i=0}^{2n_a} W_i^{c} [\boldsymbol{y}_{k}^{-, (i)} - \hat{\boldsymbol{y}}_{k}^-][\boldsymbol{y}_{k}^{-, (i)} - \hat{\boldsymbol{y}}_{k}^{-}]^\top\\
	\mathbf{P}_{qy,k} &= \sum_{i=0}^{2n_a} W_i^{c} [\boldsymbol{\mathcal{Q}}_{k}^{q-, (i)} - \hat{\boldsymbol{q}}_{k}^{-}][\boldsymbol{y}_{k}^{-, (i)} - \hat{\boldsymbol{y}}_{k}]^\top
\end{align}

Finally, the Kalman gain matrix is computed as
\begin{equation}
	\mathbf{K}_k = \mathbf{P}_{qy,k} \mathbf{P}_{yy,k}^{-1},
\end{equation}
which is used to update the posterior estimate of the latent state mean and covariance conditioned on the actual measurement $\boldsymbol{y}_k$:
\begin{align}
	\hat{\boldsymbol{q}}_k^+ &= \hat{\boldsymbol{q}}_k^- + \mathbf{K}_k \left(\boldsymbol{y}_k - \hat{\boldsymbol{y}}_k^-\right), \\
	\mathbf{P}_k^+ &= \mathbf{P}_k^- - \mathbf{K}_k \mathbf{P}_{yy,k} \mathbf{K}_k^\top.
\end{align}
This procedure is applied recursively for each control step, with measurement corrections performed only when new observations are available.

\section{Control of the 1D Kuramoto-Sivashinsky}
\label{sec:results_1DKS}
The proposed control framework is now demonstrated on the one-dimensional Kuramoto-Sivashinsky (1D KS) equation, a nonlinear partial differential equation that exhibits spatiotemporal chaos. It was first introduced as a model for laminar flame front instabilities \citep{sivashinsky_ksflame_1980, Sivashinsky_KS_1988} and chemical reaction-diffusion processes \citep{kuramoto_ksdiff_1978}. The equation is generally used as a prototypical model for chaotic dynamics, as it shares certain features characteristic of the Navier–Stokes equations \citep{Dankowicz_KS_1996}.

Let  $x(\xi, t)$  be the system state, where  $\xi$  is the spatial coordinate. The temporal evolution of the system can be described by the following equation:
\begin{equation}
    \frac{\partial x}{\partial t} 
    + x \frac{\partial x}{\partial \xi} 
    + \frac{\partial^2 x}{\partial \xi^2} 
    + \frac{\partial^4 x}{\partial \xi^4} 
    = 0
    \label{eq:1DKS}
\end{equation}
The equation includes a nonlinear convection term, a diffusion term, and a hyperdiffusion term, respectively. The 1D KS is defined on the spatial domain $\xi \in [0, L]$, with $L = 22$, and periodic boundary conditions $x(\xi, t) = x(\xi + L, t)$. The choice of $L = 22$, following \cite{Bucci_ControlChaotic_2019}, ensures that the 1D KS dynamics exhibit features similar to those observed in the Navier-Stokes equations. To enable control, a spatiotemporal forcing term $\Gamma(\xi, t)$ is added in the right-hand side of the equation, acting as the control input \citep{Bucci_ControlChaotic_2019}:
\begin{equation}
\Gamma(\xi, t) = \sum_{i=1}^{4} u^{(i)}(t) \, \frac{1}{\sqrt{2\pi\sigma_a^2}} \exp\left( -\frac{(\xi - \xi_{a}^{(i)})^2}{2\sigma_a^2} \right),
\label{eq:1DKS_control}
\end{equation}
where $\xi_{a}^{(i)} = \frac{iL}{4}$ denote the actuators locations, $\sigma_a = 0.4$ defines the width of the Gaussian support, and $u^{(i)}(t)$ is the time-dependent amplitude of the $i$-th actuator. The sensors are assumed to be equispaced along the spatial domain as follows:
\begin{equation}
    \xi_{s}^{(i)} = 1 + \frac{iL}{n_y}, \quad i = 0, \ldots, n_y-1.
    \label{eq:sensors_1DKS}
\end{equation}

The controlled 1D KS equation is integrated using a Fourier expansion of the states, a procedure enabled by the periodic boundary conditions. For this purpose, $n_x = 64$ collocation points and an integration time step of $\Delta t = 0.05$ are used. This implementation follows the approach described in \cite{Bucci_ControlChaotic_2019}, to which the reader is referred for further details.

The control objective is to stabilise the system state around its nontrivial invariant solutions, denoted as $E_1$, $E_2$, and $E_3$ \citep{Papageorgiou_chaos_ks1d_1991}. These equilibria have previously been employed as control targets in reinforcement learning applications, as reported in \cite{Bucci_ControlChaotic_2019}.

\subsection{Reduced-order surrogate model}
\begin{figure}
    \centering
    \includegraphics{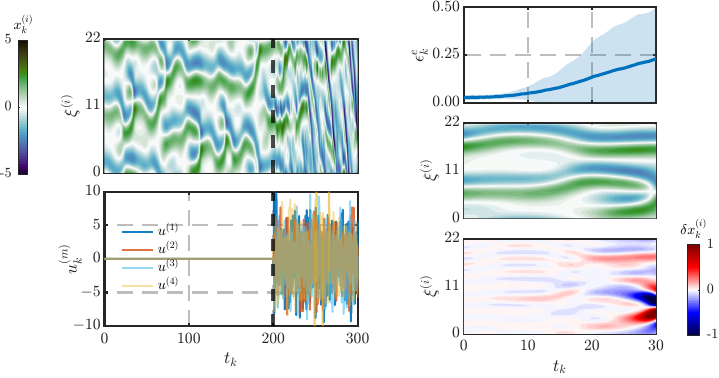}
    \caption{\justifying Left panels show the training dataset used for system identification via operator inference. The top row displays the state map, \emph{i.e.} the spatio-temporal evolution of the state, while the bottom row shows the open-loop actuation. The training dataset is shown for the first $300~t.u.$. The right panels show the model evaluation on an ensemble of $250$ trajectories. The top row presents the prediction error ($\epsilon_k^e$) over time, with the average as a solid line and the $\pm\sigma$ confidence region shaded. The middle and bottom rows show the predicted state and the prediction error for a trajectory with median ensemble error. Each prediction starts from the ground-truth state vector.}
    \label{fig:train_KS_1D}
\end{figure}

The first step in constructing a reduced-order model for the 1D KS equation is to generate a representative training dataset under both uncontrolled and controlled conditions. Specifically, a single trajectory of the full 1D KS equation is simulated over a total time of $T_{\mathrm{tr}} = 1000$ time units (t.u.), with the first $200$~t.u. corresponding to the uncontrolled system and the remaining portion subject to an open-loop control sequence. For both dataset generation and subsequent model identification and control, a time step that is an integer multiple of the numerical integration step of the 1D KS solver is used, resulting in $\Delta t_c = 0.1$~t.u. Snapshots are therefore collected every $\Delta t_c$, yielding a total of $M = 10{,}000$ snapshots in the training set. In the controlled part of the training dataset, the forcing signal for each actuator is generated as a realisation of a Gaussian white-noise process, where independent samples are drawn from a normal distribution at each time step. The time series is then smoothed in the frequency domain using a Fourier filter with cut-off $f_{co} = 1$ to eliminate high-frequency components and avoid abrupt actuation changes.
Finally, each filtered signal is normalised to have zero mean and a standard deviation of $\sigma_u = 3$. A visualisation of the training dataset, including both system states and corresponding control actions, is presented in the left column of \cref{fig:train_KS_1D}.

The training dataset is used to identify the predictive dynamical model using OpInf. First, the POD is applied to extract a low-dimensional basis. The truncation rank $r$ is selected such that the cumulative variance is $\gamma_r = 0.9999$, resulting in a reduced-order subspace of dimension $r = 14$. The reduced-order model is then identified by solving the optimisation problem in \cref{eq:OpInf_optimization}. 
A polynomial structure is imposed on the latent dynamics, including in the operator matrix in \cref{eq:op_matrix} terms up to third order in the state variables. Specifically, the model includes: a constant term $\mathbf{c}$, a linear operator $\mathbf{A}$, a quadratic operator $\mathbf{H}$, and a cubic operator $\mathbf{G}$. The dependence on the control input is modelled linearly through the operator $\mathbf{B}$, and the second-order state-input interaction is captured by the operator $\mathbf{N}$. The optimisation is performed with a regularisation coefficient $\lambda_o = 0.886$, determined via grid search by minimising the prediction error over the training dataset as done in \cite{Kramer_OpInf_2024}.

After the surrogate model is identified, its generalisability is tested on a validation dataset with the same structure as the training dataset. The model is evaluated for an ensemble of $N = 250$ trajectories with different initial conditions and open-loop actuations. The prediction performance is quantified using a normalised root-mean-square error (RMSE):
\begin{equation}
\epsilon^e_k = 
\frac{
\sqrt{\frac{1}{n_x} \sum_{i=1}^{n_x} \left( \hat{x}_k^{(i)} - x_k^{(i)} \right)^2}}
{\sqrt{\frac{1}{n_x} \sum_{i=1}^{n_x} \left( x_k^{(i)} \right)^2}},
\label{eq:error_estimation}
\end{equation}
where $x_k^{(i)}$ and $\hat{x}_k^{(i)}$ denote the $i$th component of the true and predicted state vectors at time step $t_k$. For this case, the true initial condition to perform the open-loop prediction is assumed to be known. As shown in \cref{fig:train_KS_1D}, the model provides accurate predictions even for mid- to long-term horizons, with the ensemble mean error remaining below $10\%$ for prediction windows of up to $10~t.u.$.
It should be noted that at the initial time instant, the errors are nonzero due to the low-order reconstruction losses obtained by projecting the true initial condition onto the latent space and subsequently reprojecting it into the physical space, i.e., $\hat{\boldsymbol{x}}_0 = \mathbf{\Phi}_r \mathbf{\Phi}_r^\top \boldsymbol{x}_0$. The contained ensemble sampled standard deviation of the error also suggests that the model systematically produces accurate predictions for horizons smaller than $10~t.u.$. The prediction capabilities of the model are also proved by the median error case prediction in \cref{fig:train_KS_1D}, where the prediction error map ($\delta x_k^{(i)} = \hat x_k^{(i)} - x_k^{(i)}$) is shown. Prediction errors start to accumulate for horizons longer than $20~\mathrm{t.u.}$, due to neglected terms in the dynamics and unmodeled effects arising from the truncation to a low-dimensional subspace. This phenomenon is particularly pronounced in chaotic systems such as the 1D KS equation, where long-term prediction is inherently challenging.

\subsection{State estimation via UKF}
\begin{figure}
    \centering
    \includegraphics{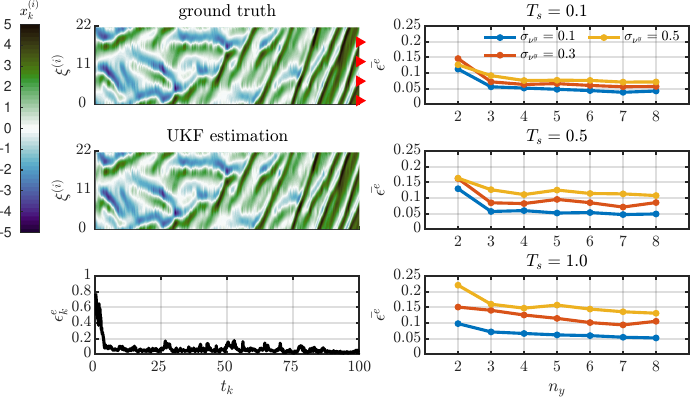}
    \caption{\justifying A posteriori Unscented Kalman Filter prediction of the state vector for controlled trajectories. Panels on the left show the ground truth of the state map and the a posteriori prediction for a case with $4$ sensors (indicated by red triangles in the top figure), measurement noise intensity $\sigma_{\nu^y} = 0.1$, and measurement sampling period $T_s = 0.1$. Panels on the right show the average prediction error $\bar{\epsilon^e}$ over the time interval $[50,\,100]$, for different configurations of noise levels, number of sensors, and update frequencies. For each case, the initial condition of the prediction is estimated using a Gaussian Process Regression based on the sensor measurements available at the initial time step.}
    \label{fig:ukf_1DKS}
\end{figure}

The UKF is now employed to estimate both the latent and full system states by combining the surrogate model predictions with available measurement data.
Figure~\ref{fig:ukf_1DKS} illustrates the application of the UKF for the state estimation problem. The full state is advanced by the OpInf model and corrected by measurements from equispaced sensors located in the domain, as given by \cref{eq:sensors_1DKS}. The sensors measure the state vector at the given location with a certain noise level, as defined in \cref{eq:measure}, and available at constant time intervals, denoted as $T_s$. The measurement noise is assumed to be white Gaussian, $\nu^y \sim \mathcal{N}(0, \sigma_{\nu^y})$, with standard deviation $\sigma_{\nu^y}$. The left column of the figure shows the case with $n_y = 4$ equispaced sensors, $\sigma_{\nu^y} = 0.1$, and measures available at every estimation time steps, leading to $T_s = \Delta t_c = 0.1$. The actuation considered during this simulation follows the same filtered Gaussian random pattern as in the training dataset, but constitutes a separate validation dataset. 

For the UKF, prior estimates of the process and measurement noise covariances, $\mathbf{Q}$ and $\mathbf{R}$, must be specified. These matrices determine the relative weighting between model predictions and measurement corrections: a larger $\mathbf{Q}$ reduces reliance on the model, while a larger $\mathbf{R}$ reduces trust in the measurements. In this work, the process noise covariance is chosen as a diagonal matrix, $\mathbf{Q} = 0.007\mathbf{I}_{n_q}$, and the measurement noise covariance is set as $\mathbf{R} = \sigma_{\nu^y}^2\mathbf{I}_{n_q}$, with the measurement variance assumed known. Both covariance matrices are manually set and kept constant during the filter execution. However, automatic procedures for selecting the covariance matrices and the Unscented Transform
hyperparameters have been proposed in the literature \citep{Scardua_TuningUKF_2017},
together with adaptive UKF formulations in which $\mathbf{Q}$ and $\mathbf{R}$ are updated online to account for variations in the noise statistics \citep{Zhang_adaptive_ukf_2022, Jiang_adaptive_ukf_2007}, but these methods are not further explored here. The unscented transformation in UKF is also configured with the parameters $\alpha = 0.1$, $\beta = 2.0$, and $\kappa = 0$. 

In the estimation process, the initial condition is assumed unknown. The model for data assimilation with UKF is initialised using an initial guess obtained via GPR based on the available partial measurements at the initial time step. By having the initial measurement $\boldsymbol{y}_0$, the GPR provides an initial state estimate $\boldsymbol{x}_0^+$ and its covariance matrix, from which the initial latent coordinate $\boldsymbol{q}_0^+$ and covariance matrix $\mathbf{P}_0^+$ are then obtained through projection. 

The left column of \cref{fig:ukf_1DKS} shows the ground truth state, the UKF posterior estimate of the full state and the corresponding estimation error, computed according to \cref{eq:error_estimation}. As expected, the initial error is large because the GPR provides only a rough first guess. However, with successive corrections from the sensors, the error converges rapidly, within $5~\mathrm{t.u.}$, to values below $0.1$, demonstrating accurate state estimation of the 1D KS system.

The estimation process for the 1D KS state is then repeated for different numbers of sensors ($n_y = 2, \ldots, 8$), noise levels ($\sigma_{\nu^y} = 0.1,\, 0.3,\, 0.5$), and UKF correction intervals ($T_s = 0.1,\, 0.5,\, 1.0~\mathrm{t.u.}$). 
As an indicator of estimation quality, the average value of the estimation error, denoted as $\bar{\epsilon^e}$, and calculated over the time window $t_k \in [50,\, 100]$, is computed for all cases using the same validation dataset. The results show a clear trend of decreasing estimation accuracy with fewer sensors, higher noise levels, and longer correction intervals.

\subsection{Set-point control}
Once the methodology for estimating the latent coordinate from sensor measurements at each time step has been established, output-feedback MPC can be applied. The initial estimate of the latent coordinate is used to evaluate the MPC cost function via model rollouts, as specified in \cref{eq:cost_function_mpc}. 

The weight matrices of the MPC cost in \cref{eq:cost_function_mpc} are selected as: $\mathbf{R}_u = 0.01\mathbf{I}_{n_u}$, and $\mathbf{R}_{\Delta u}$, $\mathbf{R}_{\Delta u} = 0.5\mathbf{I}_{n_u}$, to promote smooth control actions over time. Appropriate tuning of these matrices is crucial in problems involving multiple control targets; for instance, an automatic selection based on Bayesian optimisation is proposed in \cite{Marra_SelfTuningMpc_2024} in the framework of aerodynamic forces minimisation for a wake flow. Here, however, the control objective is simpler, \emph{i.e.} the stabilisation of the system around a target state, thus manual selection was chosen instead.
The control inputs in the MPC are constrained to lie within the bounds 
$u_k^{(i)} \in [-10,\, 10]$. For the cases considered here, the prediction and control horizons are both set to $w_p = w_c = 2~\mathrm{t.u.}$. Additionally, the terminal cost is estimated using the tail cost of a Linear Quadratic Regulator (LQR). Specifically, the OpInf model in \cref{eq:OpInf_model} is linearised around the target state $\boldsymbol{q}^*$ with zero actuation, $\boldsymbol{u} = \boldsymbol{0}_{n_u}$, and the discrete-time Riccati equation is solved to obtain the terminal cost matrix $\mathbf{Q}_f$, so that the terminal cost in the MPC formulation is $V_f(\hat{\boldsymbol{q}}) = \hat{\boldsymbol{q}}^\top \mathbf{Q}_f \hat{\boldsymbol{q}}$. This ensures that the finite-horizon MPC captures the long-term effect of the control law beyond the prediction window.

To evaluate the control performance, the normalised RMSE is considered
\begin{equation}
\epsilon^c_k = 
\frac{
\sqrt{\frac{1}{n_x} \sum_{i=1}^{n_x} \left( x_k^{(i)} - x_k^{* (i)} \right)^2}}
{\sqrt{\frac{1}{n_x} \sum_{i=1}^{n_x} \left( x_k^{* (i)} \right)^2}},
\label{eq:error_control}
\end{equation}
where $x_k^{(i)}$ and $x_k^{* (i)}$ denote the $i$th component of the true and target state vectors at time step $t_k$, respectively, and $n_x$ is the dimension of the full-order state vector. This metric provides a time-step-wise measure of deviation from the desired target state, normalised by the root-mean-square of the target.

\begin{figure}
    \centering
    \includegraphics{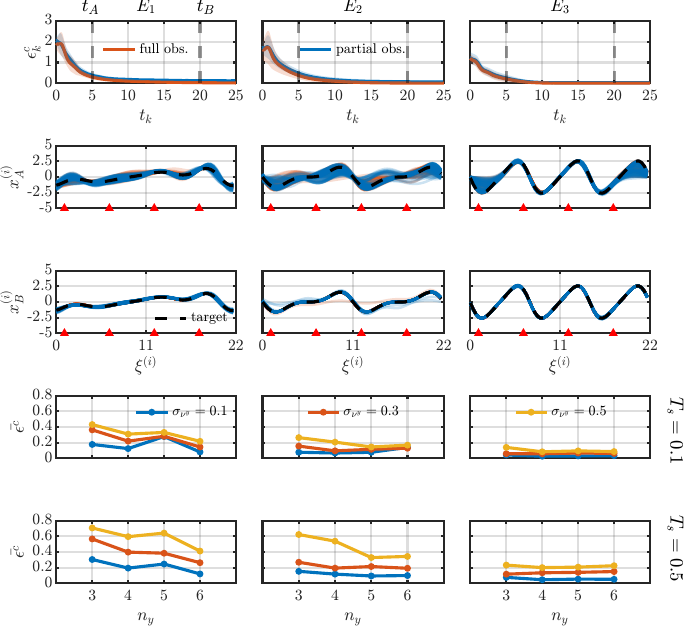}
    \caption{\justifying Results of the application of Model Predictive Control with an Unscented Kalman Filter state estimator. The control is performed around the three unstable equilibrium points of the $1$D Kuramoto-Sivashinsky system, denoted as $E_1$ (left column), $E_2$ (middle column), and $E_3$ (right column). The first three rows correspond to a configuration with $4$ sensors, measurement noise $\sigma_{\nu^y} = 0.1$, and sampling interval $T_s = 0.1$. The first row reports the control error of the true state with respect to the target (red solid line), while the blue line refers to the full-state feedback case for comparison. The second and third rows display the state evolution at two time instants, $t_A$ and $t_B$, under partial observability (red) and full observability (blue). The last two rows present the ensemble-averaged error ($\bar{\epsilon^c}$) computed over the final $5~t.u.$ of control.}
    \label{fig:control_ss_1DKS}
\end{figure}

Figure~\ref{fig:control_ss_1DKS} presents the results of the control around the three unstable invariant solutions, $E_1$, $E_2$, and $E_3$. For each case, an ensemble of $100$ control trajectories is performed from random initial conditions. It is important to note that only partial measurements of the state are available, and the initial condition is estimated from the initial sensor measurements as in \cref{fig:ukf_1DKS}. The UKF is run in a warm-up phase lasting $100~\mathrm{t.u.}$, after which control is applied for $25~\mathrm{t.u.}$.

For comparison, the figure also shows the ideal case of full observability, where the feedback consists of the complete, noise-free state of the 1D KS. A specific case with $n_y = 4$ sensors, measurement noise $\sigma_{\nu^y} = 0.1$, and sampling interval $T_s = 0.1$ is highlighted. The ensemble of states is plotted at two representative time instants, $t_A$ and $t_B$, illustrating the control transients across all trajectories as they approach the target. In this scenario, the MPC successfully drives the true state to the target in all cases, with only minor discrepancies near $E_2$ attributed to local inaccuracies of the surrogate plant model rather than state estimation errors.

Beyond this case, control is tested for various configurations of sensor number ($n_y = 3,4,5,6$), noise intensity ($\sigma_{\nu^y} = 0.1,0.3,0.5$), and sampling interval ($T_s = 0.1,0.5~\mathrm{t.u.}$). To compare the different cases, the ensemble average of the mean control error ($\bar{\epsilon}^c$) over the last $5~\mathrm{t.u.}$ of the control horizon ($t_k \in [25,30]$) is reported. The trend indicates that control performance decreases as the number of sensors decreases, or as measurement noise and sampling interval increase.

\section{Scalability of the MPC framework to high-dimensional systems}
\label{sec:results_2DKS}
This section demonstrates the extension of the proposed control framework to high-dimensional systems through its application to the two-dimensional Kuramoto-Sivashinsky (2D KS) equation. Let $\xi$ and $\eta$ denote the coordinates in the two spatial dimensions. The equation is defined on the rectangular domain $[0, 2L_\xi] \times [0, 2L_\eta]$, where $L_\xi$ and $L_\eta$ represent the half-lengths of the domain along each coordinate direction. The scalar field $x(\xi, \eta, t)$ evolves over time according to
\begin{equation}
\frac{\partial x}{\partial t} + \frac{1}{2} \left| \nabla x \right|^2 + \Delta x + \Delta^2 x = 0,
\label{eq:ks2d}
\end{equation}
where $\nabla = \left( \frac{\partial}{\partial \xi}, \, \frac{\partial}{\partial \eta} \right)^\top$ is the gradient operator, $\Delta = \frac{\partial^2}{\partial \xi^2} + \frac{\partial^2}{\partial \eta^2}$ denotes the Laplacian and $\Delta^2 = \Delta(\Delta)$ is the bi-Laplacian operator. The system is subject to periodic boundary conditions
\begin{equation}
	x(\xi, \eta, t) = x(\xi + 2L_\xi, \eta, t) = x(\xi, \eta + 2L_\eta, t),
	\label{eq:periodic_bc}
\end{equation}
and the initial condition
\begin{equation}
	x(\xi, \eta, 0) = x_0(\xi, \eta).
	\label{eq:initial_cond}
\end{equation}
The solution of the KS equation $x(\xi, \eta, t)$ exhibits spatio-temporal dynamics arising from nonlinear advection (gradient term), linear diffusion (Laplacian term) and hyper-diffusion (bi-Laplacian).

For practical reasons, a rescaled version of the 2D KS equation, as presented in \cite{Kalogirou_2DKS_2015}, is adopted through the transformation:
\begin{align}
\xi &\rightarrow \frac{L_\xi}{\pi} \xi, \quad 
\eta \rightarrow \frac{L_\eta}{\pi} \eta, \quad 
\displaystyle t \rightarrow \frac{L_\xi \, L_\eta}{\pi^2} t
\end{align}
which normalises the domain to $[0, 2\pi]\times[0, 2\pi]$. Under this rescaling, the equation becomes:
\begin{equation}
\frac{\partial x}{\partial t} + \frac{1}{2} \left| \nabla_\mu x \right|^2 + \Delta_\mu x + \mu_\xi \Delta_\mu^2 x = 0
\label{eq:2DKS}
\end{equation}
with
\begin{align}
\nabla_\mu &= \left(  \frac{\partial}{\partial \xi}, \, \frac{\mu_\eta}{\mu_\xi}  \frac{\partial}{\partial \eta}\right), \quad \Delta_\mu = \frac{\partial^2}{\partial \xi^2}
 + \frac{\mu_\eta}{\mu_\xi}  \frac{\partial^2}{\partial \eta^2},
\end{align}
and
\begin{align}
\mu_\xi &= \left( \frac{\pi}{L_\xi} \right)^2, \quad 
\mu_\eta = \left( \frac{\pi}{L_\eta} \right)^2.
\end{align}
The spatial mean, whose growth does not affect the higher modes, is defined as 
\begin{equation}
	\bar{x}(t) = \frac{1}{(2\pi)^2} \int_0^{2\pi} \int_0^{2\pi} x(\xi, \eta, t) \, d\xi \, d\eta
\end{equation}
and is subtracted during the integration to enforce a mean-zero formulation.

In addition, to enable control, similarly to the one-dimensional case, the forcing term $\Gamma(\xi, \eta, t)$ is added in the right-hand side of \cref{eq:1DKS_control}. As done similarly in \cite{Jiang_2DKSFixedPoints_2025}:
\begin{equation}
\Gamma(\xi, \eta, t) = \frac{1}{2\pi\sigma_a^2} \sum_{j=1}^{n_u} u^{(j)}(t) \, \exp\left( -\frac{(\xi - \xi_{a}^{(j)})^2 + (\eta - \eta_{a}^{(j)})^2}{2\sigma_a^2} \right),
\label{eq:forcing_term}
\end{equation}
where $u^{(j)}(t)$ is the time-dependent amplitude of the $j$th Gaussian actuator located at $(\xi_{a}^{(j)}, \eta_{a}^{(j)})$, the width is set to $\sigma_a = 0.8$ and the actuator centres are arranged on a $4\times4$ equispaced grid, yielding total of $n_u = 16$ actuators. The coordinates of the centres are $\{0,1,2,3\}\frac{L_\xi}{4}$ and $\{0,1,2,3\}\frac{L_\eta}{4}$ along the $\xi$ and $\eta$ coordinate respectively. Similarly, the sensors are assumed to be arranged in a regular grid with coordinates $\{1,\ldots,n_y\}\frac{L_\xi}{n_y+1}$ and $\{1,\ldots,n_y\}\frac{L_\eta}{n_y+1}$ along the two spatial dimensions.
In addition, to monitor the behaviour of the solution, the energy of the state is defined as
\begin{equation}
    E(t) = \int_{0}^{2\pi}\int_{0}^{2\pi} x^2(\xi, \eta, t) ~d\xi ~ d\eta.
\end{equation}

The equations are integrated using the spectral Galerkin solver \textit{Shenfun}, a partial differential equations solver provided by \cite{Mortensen_Shenfun_2018}. The equation is integrated on a computational grid of $32 \times 32$ points with an integration time step of $\Delta t = 0.0025$. This spatial and temporal resolution was found to be sufficient after comparison with two reference simulations: one using a $64 \times 64$ grid with the same integration time step, which showed solution differences of order $O(10^{-5})$, and another using a $32 \times 32$ grid with a smaller time step of $\Delta t = 0.001$, which exhibited errors of order $O(10^{-3})$.

The dynamics of the equation are determined by the selection of the scaling factor of the domain $\mu_\xi$ and $\mu_\eta$, ranging from steady solutions to chaos. A study of the behaviour of the 2D KS equation for different realisations of the scaling parameter is proposed in \cite{Kalogirou_2DKS_2015}. The parameters selected in this case are equal to $\mu_\xi = 0.3$ and $\mu_\eta = 0.16$. Under this selection, the system exhibits a modulated periodic regime, alternating between stable travelling waves and oscillatory phases with repeated, identical amplitude patterns.

A dataset is first generated, after which the OpInf procedure is applied. The time step for model identification and subsequent control of the system is set to $\Delta t_c = 0.02$. The total duration of the training dataset is chosen as $T_{\mathrm{tr}} = 3000$~t.u., yielding $M = 150{,}000$ snapshots. The dataset includes the first $500$~t.u. of the unforced solution, followed by a controlled solution with Fourier-filtered Gaussian actuations with a frequency cut-off $f_{co} = 2$. After the frequency filter, the control actions are centred to have zero mean and standard deviation $\sigma_u = 3$.

\begin{figure}

    \centering
    \includegraphics[]{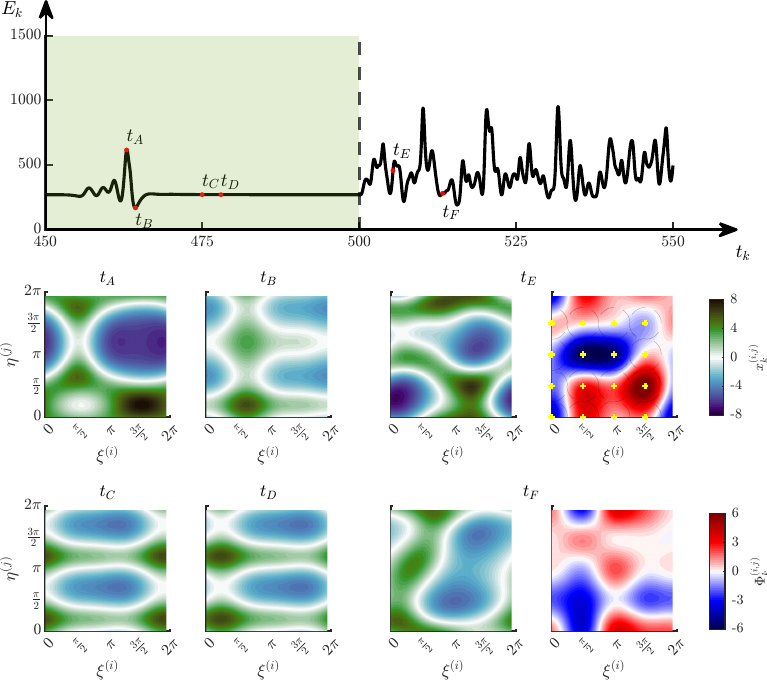}
    \caption{\justifying Representation of a subset of the training dataset used for system identification in the $2$D Kuramoto–Sivashinsky case. The central panel depicts the time evolution of the state energy over the interval $t_k \in [450,\,550]$. The green-shaded region corresponds to uncontrolled dynamics, while the unshaded region includes random, smooth actuation. The top and bottom panels display the state field (blue–green colourmap) at selected time instants within the uncontrolled ($t_A$, $t_B$, $t_C$, $t_D$) and controlled ($t_E$, $t_F$) regions. For times $t_D$ and $t_F$, the corresponding actuation maps (blue–red colourmap) are also shown. Yellow crosses indicate the actuator centres, and the contour of each actuator region corresponding to one standard deviation of the Gaussian is shown.}
    \label{fig:training_dataset_2DKS}
\end{figure}

A representation of the training dataset is included in \cref{fig:training_dataset_2DKS}, corresponding to the time interval $t_k \in [450, 550]$. The figure highlights both the free-evolution (green shaded region) and the forced segment. Snapshots of the system states at selected times during the unforced phase ($t_A$, $t_B$, $t_C$, and $t_D$) and during the forced phase ($t_E$ and $t_F$) are displayed alongside the corresponding control fields. In the uncontrolled case, intervals with nearly constant solution energy correspond to travelling waves (as seen in the snapshots at $t_C$ and $t_D$), where the solution translates across the domain at a speed determined by the wave. The figure also illustrates the oscillatory pattern that arises when the travelling wave becomes unstable. This pattern repeats periodically in time and smoothly evolves from the preceding travelling wave solution, as shown at $t_A$ and $t_B$. The figure also shows the configuration of the actuations. At time instant $t_E$, the positions of the actuator centres are indicated by yellow cross points, and the contour corresponding to one standard deviation of the Gaussian actuators is also shown.

%\subsection{State estimation and control in the two-dimensional case}

Using the training dataset described above, the OpInf method is applied to learn a surrogate model for the 2D KS system.
The number of retained modes for the low-dimensional reconstruction is determined based on the variance captured, with $\gamma_r = 0.9999$, resulting in a reduced space of dimension $r = 60$. The model as described by \cref{eq:model_deps} includes the constant term $\mathbf{c}$, the linear state operator $\mathbf{A}$, the quadratic state operator $\mathbf{H}$, and the linear input operator $\mathbf{B}$.
Regularisation for the optimisation of the operator coefficients is set to $\lambda_o = 0.1$, following the same approach used for the 1D KS case.
The resulting surrogate model, including quadratic polynomial terms, is then used in a data assimilation process where sparse, noisy measurements of the state provide feedback in the UKF framework.

\begin{figure}
    \centering
    \includegraphics{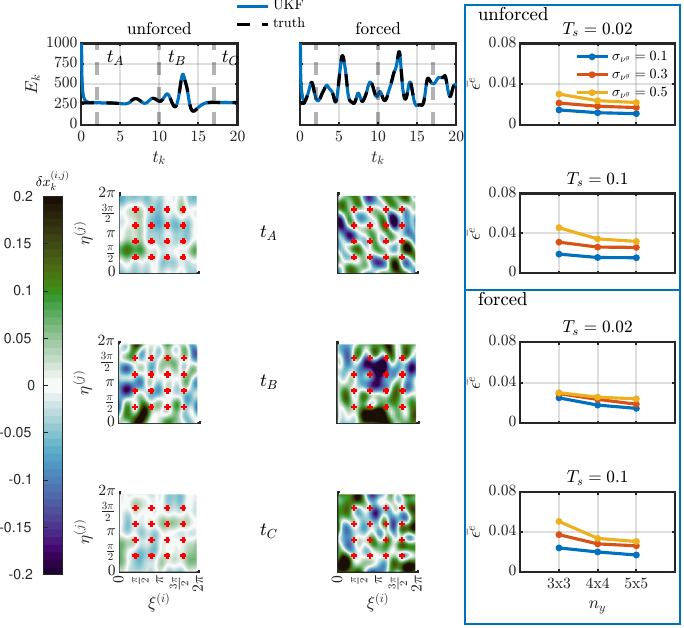}
    \caption{\justifying A posteriori UKF estimation of the state field for the $2$D Kuramoto–Sivashinsky system in controlled and uncontrolled conditions. The left panels correspond to an uncontrolled case with a $4\times4$ sensor grid, measurement noise $\sigma_{\nu^y} = 0.1$, and sampling interval $T_s = 0.1$, while the central panels show the corresponding controlled case with random smooth actuation. Red dots indicate the sensor locations for the depicted $4\times4$ configuration. For both cases, the first row presents the temporal evolution of the state energy $E_k$ in the time interval $t_k \in [0, 20]$ for the ground truth and the Unscented Kalman Filter estimate, followed by the estimation error maps at three time instants, $t_A$, $t_B$, and $t_C$. The right panels report the average estimation error ($\bar{\epsilon^e}$) over the final $10~t.u.$ of the trajectories, for different configurations of sensor grids, measurement noise levels, and sampling intervals.}
    \label{fig:ukf_KS2D}
\end{figure}

Figure~\ref{fig:ukf_KS2D} shows the results of the UKF-based state estimation for both uncontrolled and controlled scenarios.
The covariance matrices of the process and measurement noise are fixed to $\mathbf{Q} = 0.01\mathbf{I}_{n_q}$ and $\mathbf{R} = \sigma_{\nu^y}^2\mathbf{I}_{n_q}$. The unscented transformation in UKF is again configured with the parameters $\alpha = 0.1$, $\beta = 2.0$, and $\kappa = 0$.
The initial condition is assumed unknown and estimated via GPR using the available sensor measurements at the initial time.
In the left and centre columns of the figure, the UKF is tested using a regular $4\times4$ sensor grid, measurement noise level $\sigma_{\nu^y} = 0.1$, and a sampling interval of $T_s = 0.02$.
The left panels correspond to the uncontrolled case, while the central panels show the controlled scenario, using actuation patterns similar to those employed in generating the training dataset.
For both cases, the figure reports the temporal evolution of the state energy $E_k$ in a time range $t_k \in [0, 20]$, the estimation error maps at three representative time instants ($t_A$, $t_B$, $t_C$), and the locations of the sensors (red dots). The estimation error, defined in Eq.~\eqref{eq:error_estimation} and averaged over $t_k \in [10, 20]$, is evaluated for different sensor configurations (from $3\times3$ to $5\times5$ grids), noise levels ($\sigma_{\nu^y} = 0.1, 0.3, 0.5$), and sampling intervals ($T_s = 0.02, 0.1$). The curves for $\sigma_{\nu^y} = 0.1$, $0.3$, and $0.5$ are shown in blue, red, and yellow, respectively. In all tested cases, the estimation error remains below $\approx 5\%$.
Notably, the $4\times4$ sensor configuration achieves estimation accuracy comparable to the $5\times5$ grid, indicating that further increasing the number of sensors provides only marginal improvements.

The learned model is subsequently employed within the MPC framework.
The control objective is formulated as a set-point regulation problem, where the target state corresponds to the steady-state solution observed for $\mu_\xi = \mu_\eta = 0.9$.
The weighting matrices in the cost function are chosen as $\mathbf{R}_q = \mathbf{I}_{n_q}$, $\mathbf{R}_u = 0.01\mathbf{I}_{n_u}$, and $\mathbf{R}_{\Delta u} = 0.5\mathbf{I}_{n_u}$.
The control inputs are constrained within $u_k^{(i)} \in [-15,15]$, and both the prediction and control horizons are set to $w_p = w_c = 1.5$~t.u..

\begin{figure}
    \centering
    \includegraphics[]{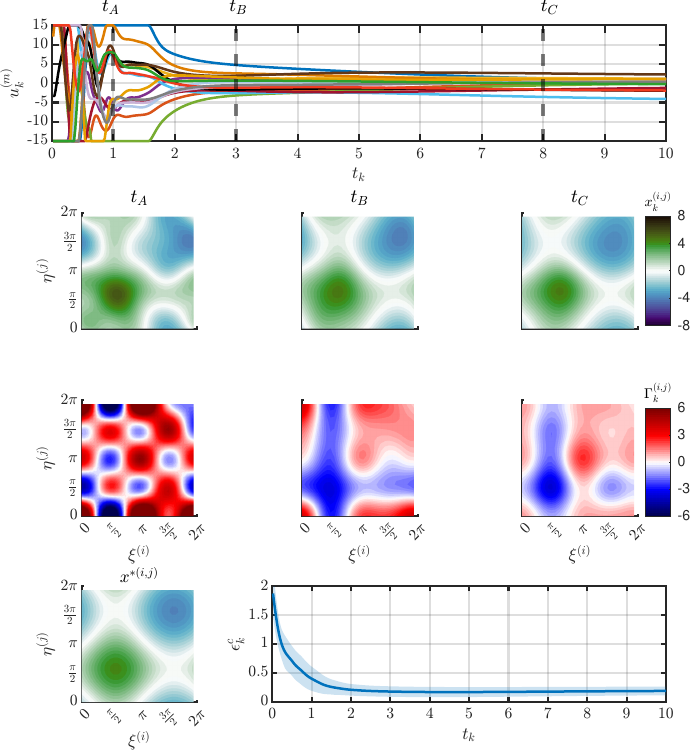}
    \caption{\justifying Control of the $2$D-KS system under full observability, where the full state is provided as control feedback. The top section illustrates the actuation history for a median-error control trajectory (top row), the corresponding state fields (second row), and the actuation maps (third row) at three different time instants, $t_A$, $t_B$, and $t_C$. The bottom panels display the target state (left) and the average control error (blue solid line) with its $\pm\sigma$ confidence region (shaded blue area) computed over an ensemble of $100$ control trajectories.}
    \label{fig:control_fo_2DKS}
\end{figure}

Before addressing the partially-observable control case, the performance of the learned surrogate model is first assessed under full-state feedback conditions.
In this configuration, the latent state is directly available to the controller at each step.
Figure~\ref{fig:control_fo_2DKS} summarises the results for an ensemble of $100$ control trajectories, each $10$~t.u. long and initialised from distinct conditions to assess generalisability.
The top panel shows the control sequence for a representative trajectory corresponding to the median control error within the ensemble.
The evolution of the $16$ actuator amplitudes (weights of the Gaussian actuators) reveals an initial coordination phase that drives the solution towards the steady-state target.
The subsequent panels display the controlled state fields and actuation maps at times $t_A$, $t_B$, and $t_C$.
The bottom panels of the figure report the target field (left) and the ensemble-averaged control error (right), as defined in Eq.~\eqref{eq:error_control}, together with its $\pm\sigma$ confidence interval. Across the ensemble, the mean control error stabilises around $19\%$, confirming that the OpInf-based model provides reliable performance for state regulation under full observability.

\begin{figure}
    \centering
    \includegraphics[]{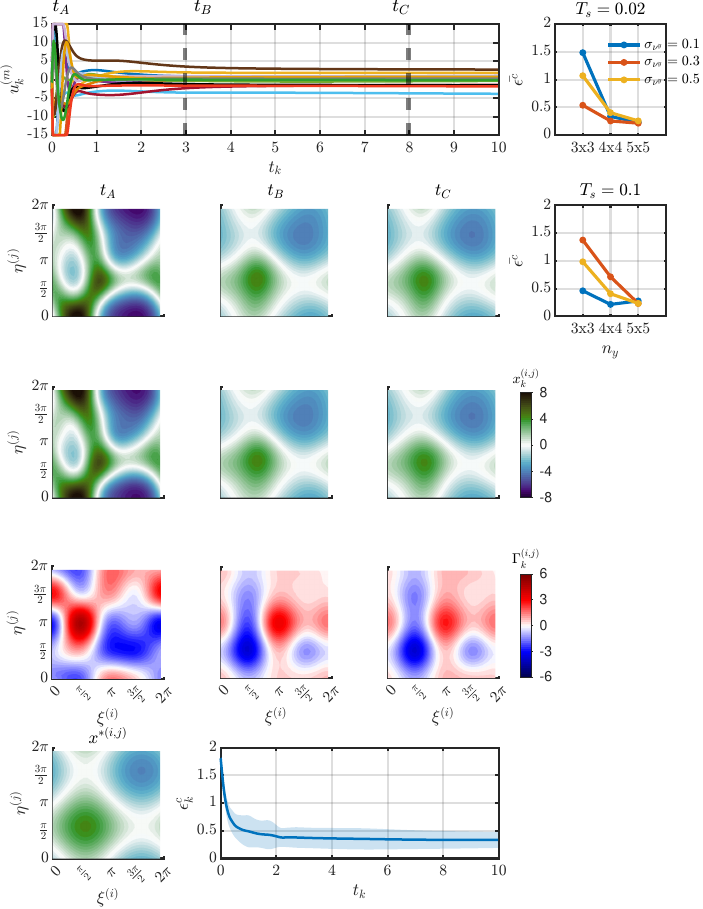}
    \caption{\justifying Control of the $2$D Kuramoto–Sivashinsky system under partial observability. The top row shows the actuation history over time for a representative median error control trajectory. The second row displays the true state at three selected control instants, $t_A$, $t_B$, and $t_C$, while the third row shows the corresponding estimated states using the Unscented Kalman Filter. The fourth row presents the MPC-computed actuation maps at the same instants. The bottom panels summarise the target state and the control error over time for an ensemble of $20$ trajectories. The ensemble-averaged estimation error ($\bar{\epsilon^c}$) over the final $2~t.u.$ of control, for different sensor numbers, noise levels, and sampling intervals, is shown in the rightmost column.}
    \label{fig:control_po_2DKS}
\end{figure}

Finally, the partially observed control scenario is presented in \cref{fig:control_po_2DKS}.
Here, the feedback to the MPC is provided by the state estimate obtained from the UKF using sparse and noisy measurements.
The study explores the effect of varying the number of sensors (from $3\times3$ to $5\times5$ grids), the measurement noise level ($\sigma_{\nu^y} = 0.1, 0.3, 0.5$), and the sampling interval ($T_s = 0.02, 0.1$). For each configuration, an ensemble of $20$ control trajectories with distinct initial conditions is simulated.
The control performance is quantified using the time-averaged error over the final $2$~t.u. of each trajectory, and averaged across the ensemble. The results show that increasing the number of sensors consistently improves both the estimation and control accuracy, with the $5\times5$ configuration outperforming the sparser grids in all tested scenarios. A representative trajectory for the case with a $4\times4$ sensor grid, $\sigma_{\nu^y} = 0.1$, and $T_s = 0.02$ is also displayed in the figure.
The top row presents the temporal evolution of the actuation history, followed by the true and UKF-estimated state fields at three control instants ($t_A$, $t_B$, $t_C$), and the corresponding actuation maps.
The rightmost panels summarise the target state, the ensemble-averaged control error, and the averaged estimation error for all sensor and noise configurations.
Overall, these results demonstrate that the combination of OpInf-based model reduction, UKF estimation, and MPC control achieves good performance even under sparse and noisy partial observations.

\section{Conclusion}
\label{sec:conclusions}
This work proposed a model predictive control framework tailored for high-dimensional nonlinear systems with complex spatio-temporal dynamics. The framework directly tackles the challenge of the scalability of model predictive control in high-dimensional systems with limited observability. The proposed approach integrates reduced-order modelling, state estimation, and control within a consistent latent-space formulation.
The predictive model is obtained via Operator Inference on a Proper Orthogonal Decomposition basis, yielding a compact, data-driven surrogate that captures the essential nonlinear behaviour of the original system. State feedback is achieved through an Unscented Kalman Filter that estimates the latent state from sparse and noisy measurements, ensuring accurate reconstruction despite limited sensing. 

The method has been validated on one- and two-dimensional Kuramoto–Sivashinsky systems, demonstrating effective closed-loop stabilisation and reliable estimation performance. These results highlight the potential of the proposed approach as a tractable and general strategy for feedback control of complex, high-dimensional systems like the ones arising in fluid dynamics.

It is thus demonstrated that integration of an unscented Kalman Filter estimator within the framework of model predictive control based on latent-space formulations enables robust kick-off of the control by progressive regularisation of the full-state estimate, and proper steering of the predictive model to avoid error divergence due to truncation of the dynamics of modelling inaccuracies in the latent space. This approach is envisioned as a solid framework applicable to fields that are severely affected by both limited observability and high dimensionality, which are ubiquitous in fluid dynamics, chemistry, biology and climate science. 
From this perspective, the proposed methodology has to be viewed as an alternative to standard control strategies, or to reinforcement learning, where partial observability is typically addressed through memory-based mechanisms such as recurrent neural networks or attention-based architectures \citep{Weissenbacher_memoryRL_2025}. Unlike memory-dependent Reinforcement Learning approaches, this framework uses an explicit state-estimation procedure, thereby avoiding issues of long-term error accumulation. This highlights the potential of latent-space, data-driven control methods as a complementary approach for controlling complex chaotic systems under limited sensing, balancing between interpretability and computational efficiency.

Future work will focus on the development of robust model predictive control extensions, quantifying modelling and estimation errors. Further investigations will target applications to more challenging fluid-dynamics configurations. Additionally, alternative latent-space representations, such as neural-network-based autoencoders, will be explored to improve model expressiveness and scalability.

\vskip6pt

\backsection[Funding]{This work is supported by the funding under ‘Orden 3789/2022, del Vicepresidente, Consejero de Educación y Universidades, por la que se convocan ayudas para la contratación de personal investigador predoctoral en formación para el año 2022’.}

\backsection[Declaration of interests]{The authors report no conflict of interest.}

\backsection[Data availability statement]{The code and data used in this work will be made available upon publication}

\backsection[Author ORCIDs]{\\
\textcolor{orcidlogocol}{\orcidlink{0000-0001-9422-2808} 0000-0001-9422-2808} - Luigi Marra;\\
\textcolor{orcidlogocol}{\orcidlink{0000-0002-0130-0545} 0000-0002-0130-0545} - Onofrio Semeraro;\\
\textcolor{orcidlogocol}{\orcidlink{0000-0003-1433-4987} 0000-0003-1433-4987} - Lionel Mathelin;\\
\textcolor{orcidlogocol}{\orcidlink{0000-0001-8537-9280} 0000-0001-8537-9280} - Andrea Meilán-Vila;\\
\textcolor{orcidlogocol}{\orcidlink{0000-0001-9025-1505} 0000-0001-9025-1505} - Stefano Discetti.}

%%% THE BIBLIOGRAPHY
%\bibliographystyle{jfm}
%\bibliography{references.bib}

%%% END OF DOCUMENT
\end{document}